\documentclass[twocolumn,showpacs,superscriptaddress,preprintnumbers,amsmath,amssymb,epsfig,floatfix,prx]{revtex4-1}
\usepackage{amsmath}
\usepackage{multirow}
\usepackage{amsfonts}
\usepackage{amssymb}
\usepackage{calligra}
\usepackage{calrsfs}
\DeclareMathAlphabet{\mathcalligra}{T1}{calligra}{l}{m}
\usepackage{epsfig}
\usepackage{graphicx}
\usepackage{dcolumn}
\usepackage{color}
\usepackage{natbib}  
\usepackage{hyperref}
\usepackage{breakurl}
\usepackage{gensymb}
\hypersetup{colorlinks=true, citecolor=blue, urlcolor=blue, linkcolor=blue}
\usepackage{bm}
\usepackage{tabularx}
\newcolumntype{L}[1]{>{\raggedright\arraybackslash}p{#1}}
\newcolumntype{C}[1]{>{\centering\arraybackslash}p{#1}}
\newcolumntype{R}[1]{>{\raggedleft\arraybackslash}p{#1}}

\begin{document}
\title{Evolution of the structural, magnetic and electronic properties of the triple perovskite Ba$_{3}$CoIr$_{2}$O$_{9}$}
\author{Charu Garg}
\affiliation{Department of Physics, Indian Institute of Science Education and Research, Dr. Homi Bhabha Road, Pune 411008, India}
\author{Deepak Roy}
\affiliation{Department of Physics, Indian Institute of Science Education and Research, Dr. Homi Bhabha Road, Pune 411008, India}
\author{Martin Lonsky}
\affiliation{Institute of Physics, Goethe-University Frankfurt, 60438 Frankfurt (M), Germany}
\author{Pascal Manuel}
\affiliation{ISIS Pulsed Neutron Source, STFC Rutherford Appleton Laboratory, Didcot, Oxfordshire OX11 0QX, United Kingdom}
\author{Antonio Cervellino}
\affiliation{Swiss Light Source, Paul Scherrer Institute, CH-5232 Villigen, Switzerland}
\author{Jens M{\"u}ller}
\affiliation{Institute of Physics, Goethe-University Frankfurt, 60438 Frankfurt (M), Germany}
\author{Mukul Kabir}
\affiliation{Department of Physics, Indian Institute of Science Education and Research, Dr. Homi Bhabha Road, Pune 411008, India}
\affiliation{Centre for Energy Science, Indian Institute of Science Education and Research, Dr. Homi Bhabha Road, Pune 411008, India}
\author{Sunil Nair}
\affiliation{Department of Physics, Indian Institute of Science Education and Research, Dr. Homi Bhabha Road, Pune 411008, India}
\affiliation{Centre for Energy Science, Indian Institute of Science Education and Research, Dr. Homi Bhabha Road, Pune 411008, India}
\date{\today}

\begin{abstract} 
We report a comprehensive investigation of the triple perovskite iridate Ba$_{3}$CoIr$_{2}$O$_{9}$. Stabilizing in the hexagonal $P6 _{3}/mmc$ symmetry at room temperature, this system transforms to a monoclinic $C2/c$ symmetry at the magnetic phase transition. On further reduction in temperature, the system partially distorts to an even lower symmetry ($P2/c$), with both these structurally disparate phases coexisting down to the lowest measured temperatures. The magnetic structure as determined from neutron diffraction data indicates a weakly canted antiferromagnetic structure, which is also supported by first-principles calculations. Theory indicates that the Ir$^{5+}$ carries a finite magnetic moment, which is also consistent with the neutron data. This suggests that the putative $J=0$ state is avoided.  Measurements of heat capacity, electrical resistance noise and dielectric susceptibility all point towards the stabilization of a highly correlated ground state in the Ba$_{3}$CoIr$_{2}$O$_{9}$ system.
\end{abstract}
\maketitle

\section{Introduction}
Due to the extensive chemical flexibility they offer, transition metal based perovskites and their variants have provided a fertile playground for exploring the complex interplay between the lattice, spin and orbital degrees of freedom. In this context, the utility of the  5$\textit{d}$ transiton metal based oxides - especially the iridates - have come to the fore very recently. Owing to the presence of 5$\textit{d}$ orbits and their inherently extended nature, these systems were expected to be relatively simple metals, devoid of strong correlation effects. On the contrary, we now know that the competing crystal field splitting, spin-orbit coupling, on-site Coulomb interaction, and Hund's coupling lead to many novel electronic and magnetic states ranging from Mott insulators to conventional long range magnetic order to quantum spin and orbital liquids in the iridates ~\citep{ annurev-conmatphys-031115-011319}. Further, the physics dramatically depends on the electronic configurations of the partially filled $t_{2g}$ orbital that leads to different magnetic ground states. In the presence of strong spin-orbit coupling, the threefold degenerate $t_{2g}$ states split into the $j_{\rm eff}$ = 3/2 quartet and the $j_{\rm eff}$ = 1/2 doublet giving rise to non-trivial magnetic solutions. The $d^1$ and $d^5$ configurations in transition metal compounds are well understood within this picture that leads to the $j_{\rm eff}$ = 3/2 and 1/2 states, respectively ~\citep{PhysRevLett.103.067205, PhysRevB.82.174440, PhysRevB.78.094403, PhysRevLett.101.076402, PhysRevLett.102.256403, Kim1329}, while the $d^2$ filling leads to non-Kramers $j_{\rm eff}$ = 2 ground state~\citep{PhysRevB.80.180409, PhysRevB.84.094420}. In contrast, a trivial spin-only $S$ = 3/2 state emerges for $d^3$ filling in a cubic environment as spin-orbit coupling gets quenched ~\citep{PhysRevLett.110.087203}. Within this description, in the $d^4$ configuration, the four electrons completely fill the $j_{\rm eff}$ = 3/2 quadruplets, when the spin-orbit coupling dominates over the Hund's coupling, and there is no local moment description. The effective spin-orbit coupling lifts this degeneracy, ultimately favouring a trivial $j_{\rm eff}$ = 0 state. In the limit when the Hund's coupling dominates over the spin-orbit coupling, the $d^4$ electrons are distributed over the three-fold degenerate $t_{2g}$ states leading to a spin $S$=1 solution. In the contrary to such naive expectations, novel magnetism is observed in the compounds with tetravalent Ru$^{4+}$ and pentavalent Ir$^{5+}$ ions with $d^4$ electrons ~\citep{PhysRevLett.112.056402, PhysRevLett.116.097205, PhysRevLett.123.017201, nphys4077, PhysRevLett.119.067201, s41467-018-06945-0}. The microscopic origin to this deviation has been debated among the gapped singlet-triplet excitonic magnetism, which can be further stabilized by the reduction in excitation energy in non-cubic crystal field ~\citep{PhysRevLett.111.197201, PhysRevLett.112.056402}, mixing of $t_{2g}-e_g$ orbitals ~\citep{PhysRevB.97.085150}, disorder ~\citep{PhysRevB.96.144423, PhysRevB.96.165108} and mixing of the $j_{\rm eff}$ = 3/2 and 1/2 states ~\citep{PhysRevB.92.121113}. Further, a quantum phase transition from the expected nonmagnetic insulator to novel magnetic state along with a rich phase diagram was predicted as a function of spin-orbit coupling, Hund's coupling and on-site Coulomb interaction~\citep{PhysRevB.91.054412}. 

\begin{figure}[!t]
\centering
\includegraphics[scale=0.6]{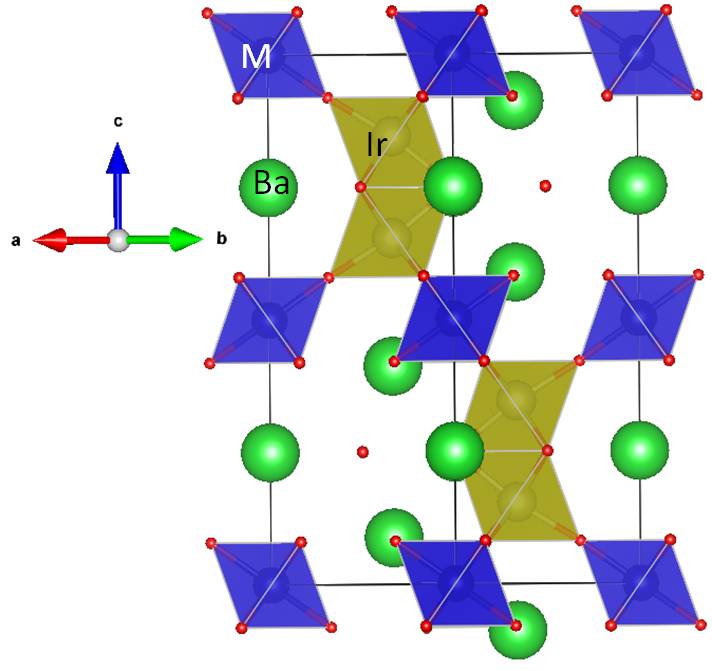}
\caption{A schematic of the crystal structure of the Ba$_{3}M$Ir$_{2}$O$_{9}$ triple perovskite. Here, blue and yellow octahedra represent cobalt and iridium sites respectively, with the Barium atoms being represented by green circles. The structure comprises of a single $M$O$_6$ octahedron sharing corners with Ir${_2}$O${_9}$ dimers, which in turn are formed by two face-sharing IrO${_6}$ octahedra.}
\label{figure1}
\end{figure}

The triple perovskite iridates of the form Ba$_3$$M$Ir${_2}$O${_9}$ (with $M$ being an alkali, alkaline earth, or transition metal) have proven to be an extremely useful structural motif in the investigation of such pentavalent Ir$^{5+}$ systems. Most of these triple perovskites are known to stabilize in the 6H-type BaTiO$_{3}$ hexagonal structure (or its distorted variants), as is depicted in Fig.~\ref{figure1}. The structure consists of IrO${_6}$ octahedra sharing a face along the crystallographic $c$ axis, with these Ir${_2}$O${_9}$ dimers being connected through corner sharing $M$O${_6}$ octahedra. The Ba ion sits in a 12 fold coordination site, as is the norm for the perovskites. The magnetic properties of the systems reported till date are primarily dictated by the multiple superexchange paths available to the Ir ions, with octahedral distortions and direct Ir-Ir exchange adding to its complexity. For instance, Ba${_3}$ZnIr${_2}$O${_9}$ was reported to exhibit a spin-orbital liquid state, with an effective moment of about 0.26 $\mu{_B}$/Ir site~\citep{PhysRevLett.116.097205}. The closely related Ba${_3}$CdIr${_2}$O${_9}$ ~\citep{PhysRevB.100.064423} and Ba${_3}$MgIr${_2}$O${_9}$ ~\citep{PhysRevB.97.064408} were also seen to exhibit no long range magnetic order down to the lowest measured temperatures. On the other hand, Ba${_3}$CaIr${_2}$O${_9}$ and Ba${_3}$SrIr${_2}$O${_9}$ systems were observed to exhibit weak-dimer-like, and ferromagnetic-like features in magnetic susceptibility respectively, though these transitions could not be identified in heat capacity measurements ~\citep{PhysRevB.97.064408}. It is to be noted that in all these systems, the $MO{_6}$ octahedra is magnetically inert. Insertion of a magnetically active ion at this site would add to the complexity of these triple perovskites, by the inclusion of additional superexchange paths which would be expected to promote long range magnetic order.

Interestingly, there has been no systematic investigation of systems where the Ir$^{5+}$ ions interact with each other and magnetic $M$ ions within the framework of the triple perovskite structure. Here we report on the structural, magnetic and electronic properties of Ba$_{3}$CoIr$_{2}$O$_{9}$ - where these conditions are met, as the B-site is shared by pentavalent Ir$^{5+}$-$d^4$ and Co$^{2+}$-$d^7$ ions. This system is seen to exhibit a magneto-structural transition at 107 K - the highest known among all the triple perovskite iridates - with the crystallographic symmetry being lowered from the high temperature hexagonal ($P6_{3}/mmc$) to a monoclinically distorted ($C2/c$) one. On further reduction in temperature, we observe the stabilization of another monoclinic phase with even lower symmetry ($P2/c$), with both these  phases coexisting down to the lowest measured temperatures. The first-principles calculations, including the spin-orbit coupling within the GGA + $U$ + SOC formalism, predicts a weakly b-canted antiferromagnetic (AFM) structure (with the main AFM component along c) as the ground state that agrees with the experimental magnetic structure derived from the neutron diffraction. Measurements of the resistivity, specific heat, dielectric susceptibility, and electronic noise all point towards the stabilization of a highly correlated ground state, with unconventional magnetism, spin-orbit coupling and structural and electronic phase co-existence being critical ingredients. 

\section{Methods}
Polycrystalline specimens of Ba$_{3}$CoIr$_{2}$O$_{9}$ were synthesized by the standard solid state reaction technique. Stoichiometric amounts of high purity BaCO$_{3}$, Co$_{3}$O$_{4}$ and IrO$_{2}$ were thoroughly ground using a mortar and pestle to get homogeneous mixtures, which were then fired in air at 950$\degree$ C for 24 hrs, followed by a heat treatment at 1200$\degree$ C for 48 hrs. These powders were then cold pressed and heat treated in air repeatedly at 1200 $\degree$C till no change in the powder X-Ray Diffraction (XRD) peak shape was observed . The XRD patterns were measured using a Bruker D8 Advance diffractometer with a Cu K$_{\alpha}$ source. Temperature dependent synchrotron XRD measurements were performed using the Materials Science (MS) X04SA beam line (wavelength 0.5653$\lambda$)  at the Swiss Light Source (SLS, PSI Switzerland)~\cite{SLS}. The powder sample was filled in a 0.3mm capillary and the experiments were carried out in the temperature range 4.2K-295K. Temperature dependent powder neutron diffraction measurements were carried out using the time-of-flight WISH diffractometer at the ISIS neutron facility~\citep{ISIS}. Crystal and magnetic structural details were analyzed by the Rietveld method using the FullProf refinement program~\cite{Fullprof}. The structures shown in the manuscript are drawn using Vesta~\cite{vesta}. Elemental compositions were reconfirmed by using an energy dispersive X-Ray spectrometer (Zeiss Ultra Plus). The X-Ray photoelectron spectroscopy (XPS) measurements were carried out on a laboratory setup (K Alpha+ model, ThermoFisher Scientific Instruments). Magnetization and physical property measurements were performed using a Quantum Design (MPMS-XL) SQUID magnetometer and Physical Property Measurement System (PPMS) respectively. Temperature dependent dielectric measurements were performed in the standard parallel plate geometry, using a NOVOCONTROL (Alpha-A) High Performance Frequency Analyzer. Measurements were typically done using an excitation ac signal of 1V at frequencies varying from $2\,$kHz to $4\,$MHz. Low-frequency fluctuation spectroscopy measurements were carried out in a continuous-flow cryostat with a variable temperature insert. During the measurements, the fluctuating voltage signal is amplified and processed by a spectrum analyzer yielding the voltage noise power spectral density (PSD) $S_V(\omega)$ defined by 
\begin{equation}
S_V(\omega)=2\lim\limits_{T \to \infty}\frac{1}{T}\left| \int_{-T/2}^{T/2} \mathrm{e}^{i\omega t}\ \delta V(t)\ \mathrm{d}t\ \right| ^2,
\end{equation}
where $\delta V(t)$ represents the fluctuating voltage drop across the sample and $\omega=2\pi f$ the angular frequency. In this study, noise spectroscopy was conducted in a five-terminal ac setup, where the sample is placed in a bridge circuit in order to suppress the constant dc voltage offset and to minimize external perturbations~\cite{acnoise}. Further details about the fluctuation spectroscopy technique can be found elsewhere~\cite{Mueller2011}. 

To get additional insight into the electronic and magnetic ground states, we performed first-principles density functional theory calculations~\citep{PhysRev.136.B864,PhysRev.140.A1133}, as realized in the VASP code \citep{PhysRevB.48.13115,PhysRevB.54.11169,PhysRevB.59.1758,doi:10.1063/1.2187006,doi:10.1063/1.2403866}. The wavefunctions are described within projector augmented wave formalism with 550 eV cutoff for the kinetic energy~\citep{PhysRevB.50.17953}. The exchange-correlation energy was described within the Perdew-Burke-Ernzerhof functional form of generalized gradient approximation ~\citep{PhysRevLett.77.3865}, and the spin-orbit coupling was considered. The reciprocal space integration was carried with a 7$\times$4$\times$3 Monkhorst-Pack $k$-mesh. The correlation effect was corrected using the Hubbard like on-site Coulomb interaction $U$ parameter, which was set to 3 and 2.4 eV for Co-3$d$ and Ir-5$d$ electrons, respectively,  within the rotationally invariant Dudarev's approach~\citep{PhysRevB.57.1505}. The experimental lattice parameters derived from the synchrotron data at 80 K were used in the theoretical calculations, and various initial magnetic configurations were considered and  optimized. 

\begin{figure}
\centering
\includegraphics[scale=0.37]{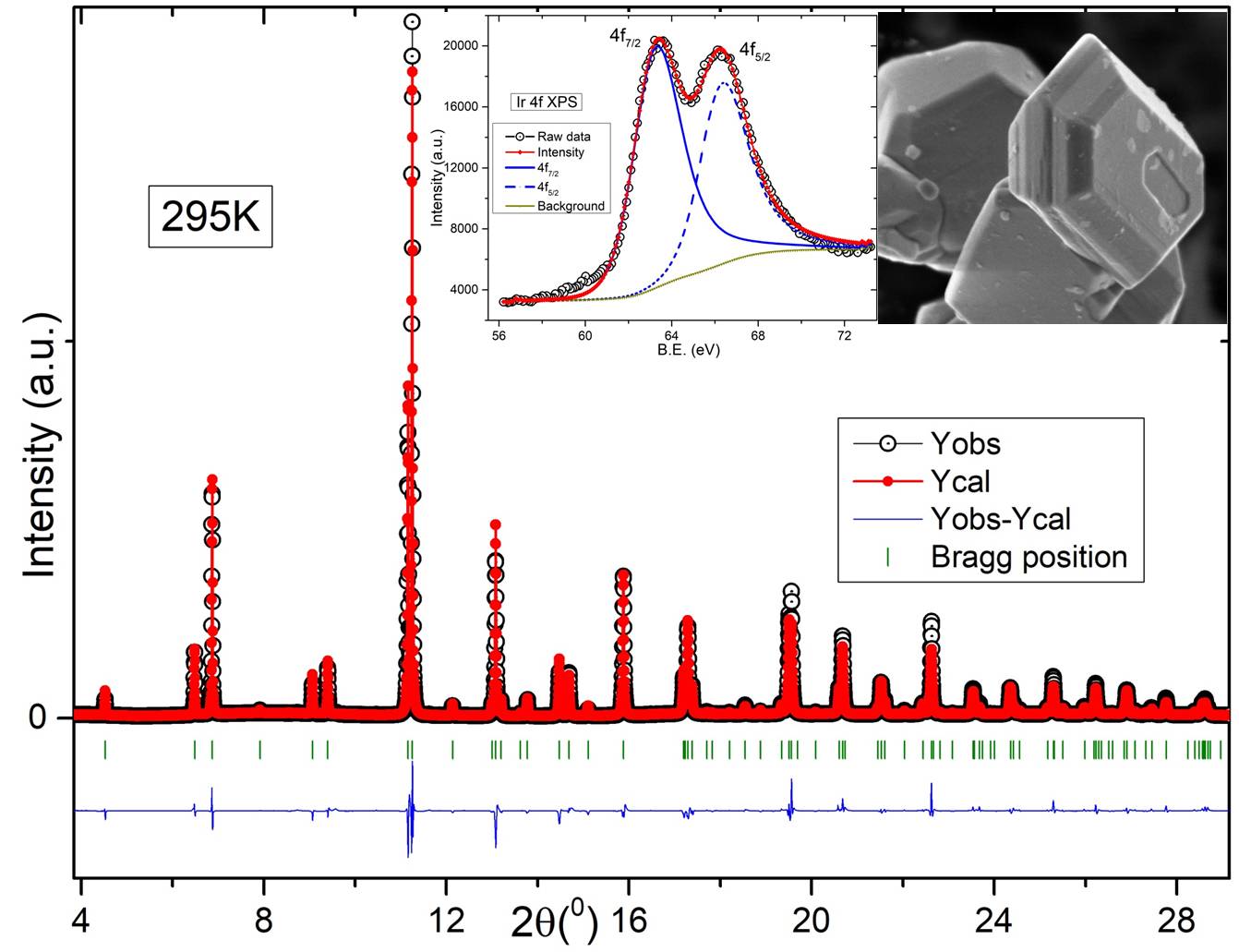}
 \caption{Main Panel: Rietveld refinement of synchrotron powder diffraction pattern of Ba$_{3}$CoIr$_{2}$O$_{9}$ at 295 K. The compound crystallizes in a 6H-hexagonal perovskite with space group (P6$_{3}$/mmc). The calculated and observed diffraction profiles are shown as red and black markers respectively. The vertical green lines indicates the calculated Bragg positions. The blue line at the bottom is the difference between observed and calculated intensities. Inset: SEM images of the system exhibiting clear hexagonal facets s shown on the right. The XPS fit for the Ir $4f$ core level is depicted on the left. Black circles are the data points and the red line is the fit. The solid and dashed blue lines depicts the deconvolution of the $4f_{7/2}$ and $4f_{5/2}$ contributions respectively.}
\label{fig2} 
\end{figure}

\begin{table}[!t]
\centering
\caption{Structural parameters of Ba$_{3}$CoIr$_{2}$O$_{9}$ at 295 K.}
\begin{tabular}{C{1.5cm} C{1.5cm}  C{1.5cm}  C{1.3cm} C{1.7cm}}
\hline
\hline
\multicolumn{5}{c}{Space group $P6_{3}/mmc$ (No. 194)} \\
\multicolumn{5}{c}{$a=b=5.7639(1)$ \AA, $c=14.2949(3)$ \AA, $\alpha=\beta=90^{\circ}$, $\gamma$=120$^{\circ}$} \\
\hline
Atom & $x$  & $y$ & $z$ & occupancy\\ \hline
    Ba1 & 0.3333 & 0.6666 & 0.9120(1) & 1\\ 
    Ba2 & 0.0000 & 0.0000 & 0.2500 & 1 \\
    Co & 0.0000 & 0.0000 & 0.0000 & 1  \\
    Ir & 0.3333 & 0.6666 & 0.1540(1) & 1\\
    O1 & 0.5004(14) & 1.0007(28) & 0.2500 & 1\\
    O2 & 0.1619(9) & 0.3238(18) & 0.4196(2) & 1\\
\hline
\end{tabular}
\label{tab:table1}
\end{table}

\section{Results and Discussions}
The room temperature synchrotron XRD data and its Rietveld refinement is depicted in Fig~\ref{fig2}. Ba$_{3}$CoIr$_{2}$O$_{9}$ is seen to crystallize in an aristotype 6H-type BaTiO$_{3}$ hexagonal structure $P6_{3}/mmc$ (space group ($\# 194$)), consistent with the previously reported triple perovskites~\cite{Ru1,hexa}. This structure comprises of IrO$_{6}$ octahedra sharing a face along the crystallographic $c$ axis, with the Ir$_2$O$_9$ dimers connected by corner sharing CoO$_{6}$ octahedra, and Ba occupying the $12$ fold coordination site. The lattice parameters obtained from the fit are $a = b = 5.7639$ \AA, $c = 14.2949$ \AA\ and $\alpha$=$\beta$=90$\degree$, $\gamma$= 120$\degree$. The resultant structural parameters from the Rietveld refinement of synchrotron data are listed in Table~\ref{tab:table1}. The chemical composition was confirmed with the help of EDX analysis, and the Ba:Co:Ir ratio was inferred to be 3.02:1:1.95. The inset of Fig.~\ref{fig2} shows the scanning electron image of the system exhibiting clear hexagonal facets. 
	
Since both cobalt and iridium can stabilize in multiple valence states, it is imperative to ascertain the valence of these ions, which in turn would critically influence the electronic and magnetic properties. Our XPS measurements of the Ir 4f core level is shown in the inset of Fig~\ref{fig2}. The best result was obtained by using only a single orbit spin doublet, and the binding energy position and gap thus obtained confirms the presence of only Ir$^{5+}$ (and by corollary Co$^{2+}$) in this system. In addition, the charge state of Ir in hexagonal lattices can also be deduced by the Ir-Ir bond distance~\citep{IrIr}, as it is reported to increase with the oxidation state of iridium, with the electrostatic repulsion resulting in the bond distances of the order of 2.55, 2.65 and 2.75 \AA\ for Ir$^{4+}$, Ir$^{4.5+}$ and Ir$^{5+} $ respectively. The obtained value of the Ir-Ir distance (=2.75 \AA) from our room temperature Rietveld refinement fit is in good agreement with the $+5$ state in iridium.

\begin{figure}
\centering
\includegraphics[scale=0.29]{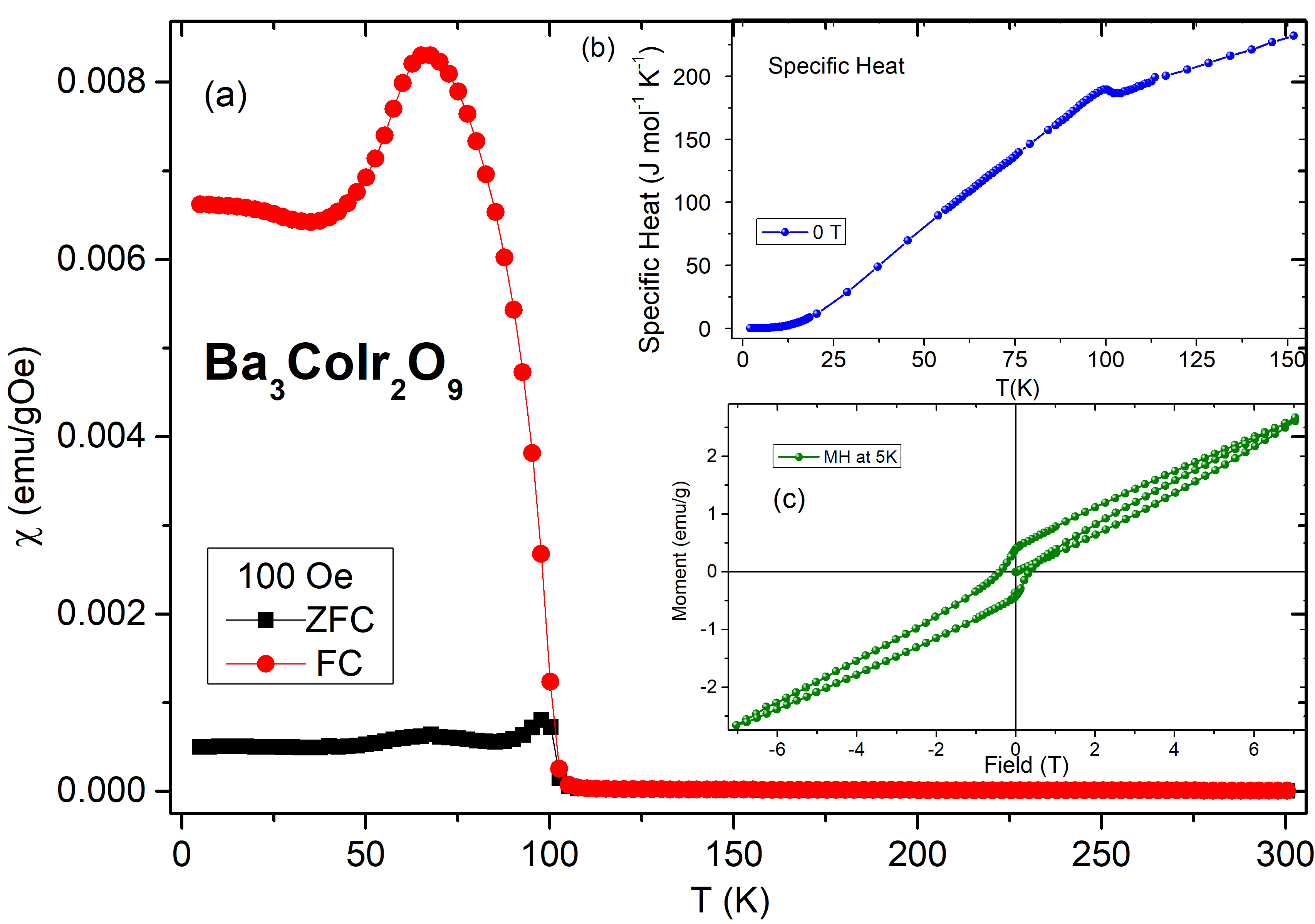}
\caption{(a) Temperature dependent magnetization of Ba$_{3}$CoIr$_{2}$O$_{9}$ as measured in the zero field cooled (ZFC) and field cooled (FC) protocols with an applied magnetic field of $100\,$Oe. (b) depicts the heat capacity as measured for this system, with a peak characterizing the magnetic transition. (c) depicts the MH isotherm as measured at $5\,$K, with a finite opening of the loop indicating a finite ferromagnetic contribution to the otherwise antiferromagnetic system.}
\label{variation}
\end{figure}

The temperature dependence of the DC magnetic susceptibility as measured in Ba$_{3}$CoIr$_{2}$O$_{9}$ is shown in Fig.~\ref{variation}, where a pronounced irreversibility between ZFC-FC curves was seen when the system orders magnetically at 107 K. This is in agreement with a solitary prior report of this system~\cite{battle} where the  magnetization was reported, although the nature of the transition remained unclear. This transition  is also evident in the measurements of the specific heat (inset of Fig.~\ref{variation}), reflecting a large change in entropy. A Curie Weiss fit to our magnetization data gives a Curie-Weiss temperature ($\theta_{\rm CW}$) of 6.35 K, and an effective magnetic moment of 4.72 $\mu_{B}$/f.u. respectively. The small positive value of $\theta_{\rm CW}$ is clearly suggestive of mixed ferro-antiferromagnetic interactions. This is also confirmed by the MH isotherm at 5K [Fig.~\ref{variation}(c)], where a non-saturating magnetization, along with a small loop opening is observed - typical of ferri-, or weak ferromagnetic systems.  

The effective  magnetic moment of 4.72 $\mu_{B}$/f.u obtained by the Curie-Weiss fit can be accounted for by considering Co$^{2+}$($d^7$) in the high spin state ($S=3/2$) and iridium in the Ir$^{5+}$($d^4$) state. The spin only contribution from Co$^{2+}$ is 3.88 $\mu_{B}$, implying that the contribution from the Ir$^{5+}$($d^4$) is 0.41 $\mu_{B}$. Though Ir$^{5+}$, by virtue of its electronic configuration (5$\textit{d}^{4}$) and under the influence of strong spin orbit coupling (t$_{2g}^{4}$ e$_{g}^{0}$) is expected to exhibit a $J=0$ non-magnetic state, our data indicates a contribution from both the magnetic entities. This breakdown of the $J=0$ state has been observed previously in systems like Ba$_{3}$ZnIr$_{2}$O$_{9}$~\cite{PhysRevLett.116.097205} and  Ba$_{3}$MgIr$_{2}$O$_{9}$~\cite{PhysRevB.97.064408} where the Ir moment was reported to be 0.2 $\mu_{B}$ and 0.5-0.6 $\mu_{B}$ respectively. A defining feature of these triple perovskites is the presence of Ir$_{2}$O$_{9}$ dimers in the backdrop of strong spin orbit coupling. It has been suggested that inter and intra dimer interactions are strongly influenced by the local distortions creating complex Ir-O-Ir pathways leading to the breaking of degenerate $t{_{2g}}$ levels. Under the influence of strong SO coupling, these states reform into mixed $J$ states, which in turn facilitate hopping induced superexchange interactions that result in a finite magnetic moment at the Ir site. 

\begin{figure}
\centering
\includegraphics[scale=0.35]{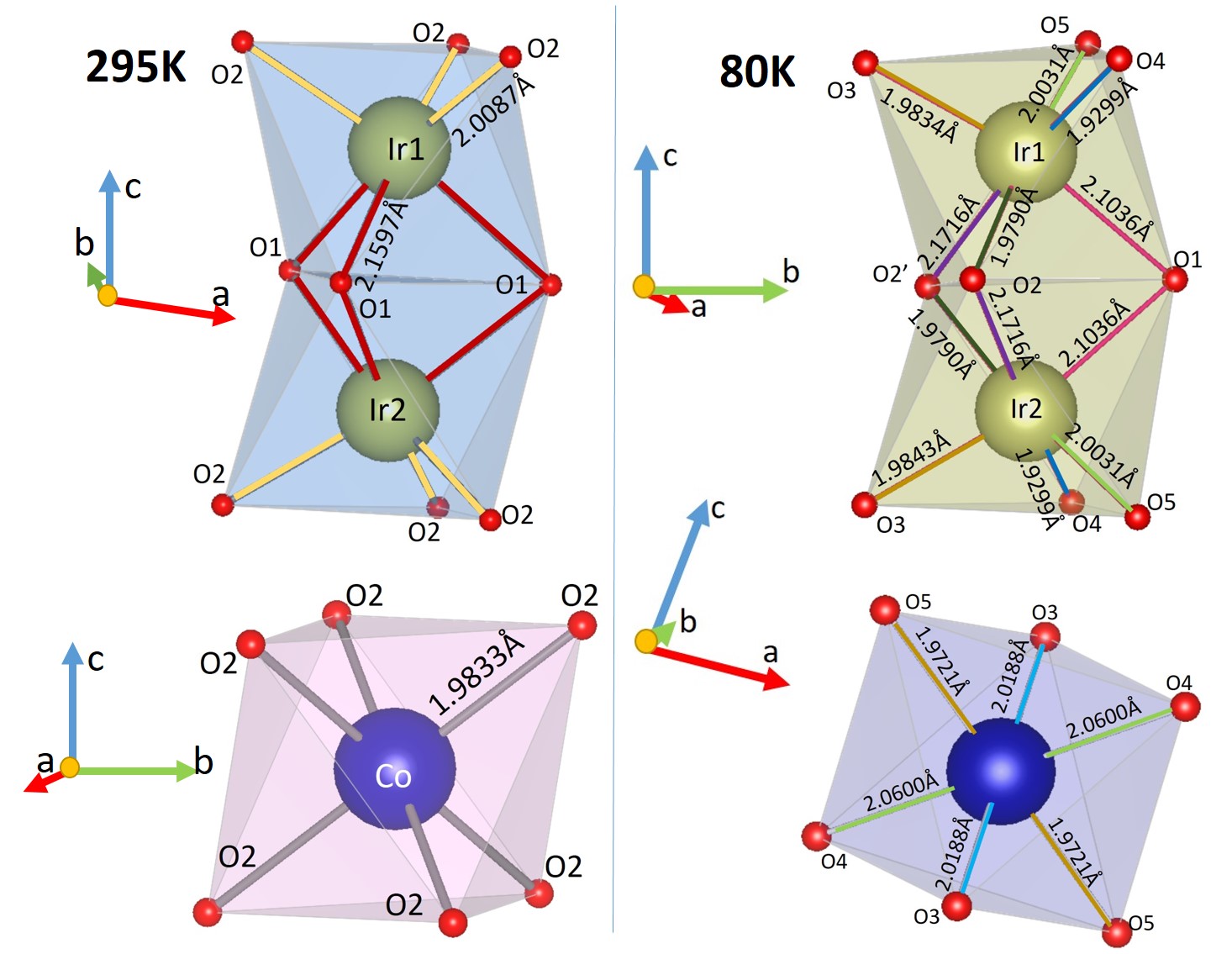}
\caption{Schematic representation of Ir dimers and Cobalt octahedras at 295 K (left) and 80 K (right). The cobalt octahedra has only one Co-O bond at high temperatures, which  splits into 3 pairs of Co-O bonds at $80\,$K. Similarly, Ir has 2 different bonds associated with Ir face sharing and corner sharing O atoms in the high symmetry structure, which transforms into 6 different pairs of Ir-O bonds below the magnetostructural transition. The bond lengths with equal magnitudes are represented with the same color for the sake of clarity.}
\label{figure4}
\end{figure}

\begin{figure}
\centering
\includegraphics[scale=0.33]{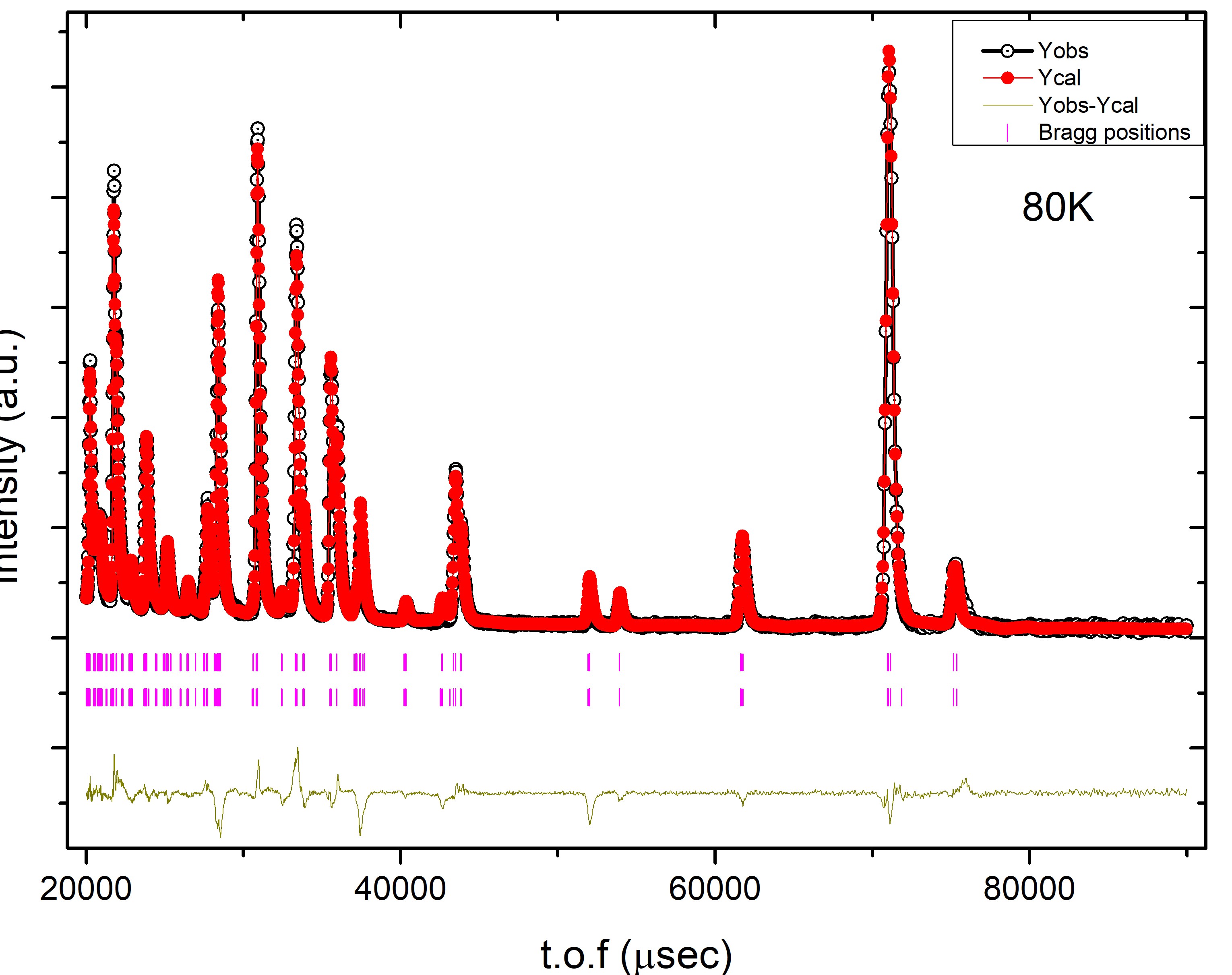}
\caption{Neutron powder diffraction pattern of Ba$_{3}$CoIr$_{2}$O$_{9}$ measured at 80\,K using the WISH diffractometer.  The calculated and observed diffraction profiles are shown in red and black respectively. The upper and lower vertical pink marks  represent the positions of nuclear and magnetic reﬂections respectively, and the green line at the bottom is the difference between observed and calculated intensities. The fit was for Ireps $\Gamma_{1}$, and the symmetry operators and magnetic basis vectors are (1) x,y,z [u,v,w], (2) -x,y,-z+1/2 [-u,v,-w], (3) -x,-y,-z [u,v,w], and (4) x,-y,z+1/2 [-u,v,-w] for Iridium at 8($\textit{f}$) and (1) and (2) for more symmetric site 4($\textit{a}$) for Cobalt. Here, u, v, and w indicate the components of the magnetic moment.}
\label{figure5}
\end{figure}

It is known that the monoclinically distorted Ba$_{3}$BiIr$_{2}$O$_{9}$ system exhibits a giant magnetoelastic transition - the largest reported amongst all $5d$ oxides - at $\sim$ 75 K. Interestingly, this transition was not accompanied by a change in space group symmetry, but was characterized by a 4$\%$ increase in the Ir-Ir bond distance and 1$\%$ negative thermal volume expansion. It was proposed that the transition occurred due to the existence of an electronically unstable (nominal) Bi${^{4+}}$ valence state, which then correspondingly influences the valency of Ir~\citep{BaBi,BaBi2,BaBi3}. However, the Bi based system is an exception, and most triple perovskite iridates -especially where Ir stabilizes in the 5$\textit{d}^{4}$ configuration - do not exhibit long range magnetic order, or at best exhibit weak dimer-like magnetism. In other triple perovskite iridates with mixed valent Ir ions, low magnetic ordering temperatures have been reported, though there are no reports of magneto-structural transitions in any of them~\cite{PhysRevB.97.064408,PhysRevLett.116.097205,Ba3LiNaK,IrIr}. Interestingly, the onset of magnetic order in Ba$_{3}$CoIr$_{2}$O$_{9}$ is accompanied by a structural transition, and the system is observed to distort from a 6H hexagonal structure ($P6{_3}/mmc$) to a monoclinic phase ($C2/c$). This leads to a doubling of the crystallographic unit cell, and the new lattice parameters are $a=5.7368$ \AA, $b=9.9833$ \AA, $c=14.2686$ \AA, with $\alpha$=$\gamma$=90$\degree$ and $\beta$= 90.0663$\degree$. As a consequence, both the Ir and Co octahedra are now highly distorted. Fig.~\ref{figure4} shows the comparison of the octahedra on either side of this magneto-structural transition. The bond-lengths with equal magnitudes are assigned same colors for the sake of clarity.  In the case of cobalt, a highly symmetric octahedron with only one Co-O bond length in the high $T$ phase transforms into $3$ pairs of Co-O bonds, with a maximum of $3.87\%$ change in the bond length below the transition. On the other hand, the iridium octahedron distorts from three Ir-O1 and Ir-O2 bonds above $T_{N}$ to six different pairs of Ir-O bonds below the transition. We note that this kind of a magneto-structural transition has not been previously observed in any Ir-based triple perovskite till date, and the Ba$_{3}$CoIr$_{2}$O$_{9}$ system exhibits the highest magneto-structural transition temperature reported for any Ir based triple perovskite.

\begin{figure}
\centering
\includegraphics[scale=0.35]{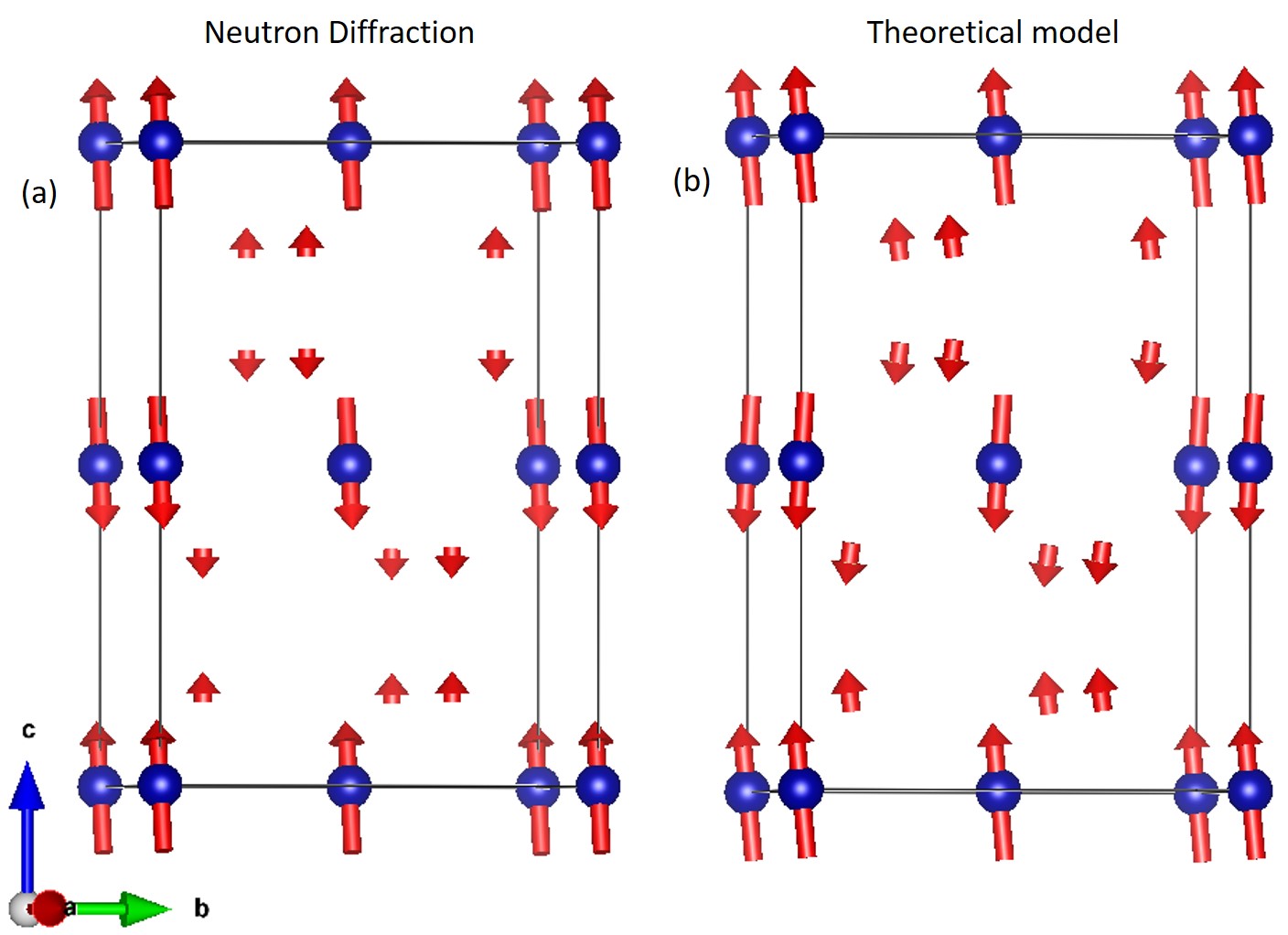}
\caption{Schematic of the magnetic structure of the monoclinic (C2/c) phase at $80\,$K from (a) the Rietveld analysis of neutron diffraction data, and (b) using first-principles GGA  + $U$ + SOC calculations. Here, the blue solid circles and the small red arrows represent cobalt and iridium atoms respectively.}
\label{figure6}
\end{figure}


The multiple Ir-O-Ir and Co-O-Ir bond angles (and bond lengths) formed as a consequence of this monoclinic distortion creates multiple exchange paths that influence the magnetic and electronic structure. Since the Co-Co distance is very large (7.14 \AA), the two primary magnetic interactions are between the Ir and Co edge sharing octahedra and amongst the Ir$_{2}$O$_{9}$ dimers. To get an insight into the magnetic structure, we performed powder neutron diffraction measurements at 80\,K (Fig.~\ref{figure5}).  The parent structure used for the symmetry analysis was $C2/c$ (Space  group: 15). Our neutron diffraction measurements indicate that the nuclear and magnetic peaks appear at the same Bragg position, suggesting that the propagation vector of the magnetic structure is $\textbf{K}$=(0,0,0) which was also confirmed by using the K search program in FullProf~\cite{Fullprof}. We performed the symmetry analysis using BasIreps to find the possible magnetic structure based on this propagation vector. Out of the multiple solutions, the two indicating zero moment on Cobalt were naturally ruled out, and we converged to two different magnetic representations, $\Gamma_{1}$and $\Gamma_{3}$, with almost similar goodness of fit ($R_{wp}/R_{e}$$\approx$ 3.3 and 3.5 respectively). Both the magnetic models showed antiferromagnetically coupled Ir dimers, although they suggest FM and AFM Co/Ir interactions along different crystallographic directions.

The first-principles calculations, including the spin-orbit coupling within the GGA + $U$ + SOC formalism, predicts a weakly b-canted AFM structure as the ground state that agrees well with the former magnetic structure derived from the neutron diffraction at 80 K (Fig.~\ref{figure6}). The spin canting is relatively more prominent in Ir sites than the Co moments. The intra-dimer Ir$-$Ir interaction is antiferromagnetic, which is similar to other iridates and ruthenates~\citep{PhysRevLett.116.097205, PhysRevLett.123.017201, PhysRevB.85.041201, PhysRevB.97.064408}, while the in-plane Co spins interact ferromagnetically. The Ir$-$Co interaction is found to be ferromagnetic, making the inter-dimer Ir$-$Ir coupling ferromagnetic and forcing antiferromagnetic alignment between the Co-planes. We also have considered several other magnetic structures; however, all configurations converged back to the ground state. The constrained collinear calculations reveal that the non-collinearity is rather weak as the commensurate collinear spin configuration lies only 1.75 meV/f.u. higher, where all the spins align along the $c$-direction, and resembles with that of a previously reported ruthenate, Ba$_3$NiRu$_2$O$_9$~\citep{LIGHTFOOT1990174}. In the present case, both electron correlation and SOC play important roles in determining the correct magnetic structure as the GGA + $U$ calculations indicate unphysical FM ground state, and this will be discussed later.    

In the non-collinear magnetic ground state, the net moment along the $c$-axis is zero due to completely compensated AFM coupling, while the (small) spin components along the $b$-direction align parallely giving rise to weak ferromagnetism. This picture agrees well with the current experimental observation that the magnetic structure is predominantly AFM with superimposed weak ferromagnetism. The obtained magnetic moment from neutron diffraction for cobalt is 2.45 $\pm{0.19}\mu_{B}$, which is close to the moment obtained (2.84 $\mu_{B}$) for the ruthenate analogue Ba$_{3}$CoRu$_{2}$O$_{9}$~\cite{PhysRevB.85.041201}.   The obtained value of moment for Co$^{2+}$ is lower than the spin only value for high spin state of cobalt (3.88 $\mu_{B}$). This could be due to hybridization effects and covalency effects as has been speculated earlier for the ruthenate analogue and the cobalt based double perovskite Sr$_2$CoWO$_6$~\citep{viola2003structure,Ru2}. An unrestrained fit of the neutron diffraction data suggests that Ir${^{5+}}$ has a finite moment value (0.57 $\pm{0.12}\mu_{B}$) which is close to the value of  0.42 $\mu_{B}$ obtained from our Curie-Weiss fit. With iridum being a strong absorber of neutrons, it is difficult to unambiguously resolve the details of the magnetic structure using the neutron data alone, and hence we rely on our first principles density functional theory calculations to substantiate our key inferences, as described below. 

The XPS data reveals Ir$^{5+}$-$d^4$ and Co$^{2+}$-$d^7$ electronic configurations, and the first-principles GGA + $U$ + SOC  calculations predict 0.86 and 2.76 $\mu_B$ local moments at the Ir and Co sites, respectively. The results are consistent with the present experimental results and previous reports on iridates~\citep{PhysRevLett.116.097205, PhysRevLett.123.017201, PhysRevB.97.064408}. Our calculations show that forcing the iridium moment to zero increases the energy by 200 meV/f.u., indicating that the ground state must have non-zero iridium moment. The corresponding orbital moments of 0.29 and 0.04 $\mu_B$ at the Ir and Co sites indicate strong spin-orbit coupling. Within the GGA + $U$ + SOC calculations, the constrained collinear moments along the $c$-direction are very similar ($\mu_{\rm Ir}$ = 0.85 and $\mu_{\rm Co}$ = 2.76 $\mu_B$) to the noncollinear moments, further indicating weak noncollinearity. This is consistent in the low value of ferromagnetic component (0.47 $\pm{0.21}\mu_B$) estimated from our neutron diffraction results. The Co$^{2+}$-$d^7$ ions stabilize in a high-spin $S=$ 3/2, $t_{2g}^5e_{g}^2$  state due to the low crystal field splitting $\Delta_{\rm Co} \sim$ 1 eV and high spin pairing energy of 2.3 eV in such compounds~\citep{LIGHTFOOT1990174, PhysRevB.58.10315, PhysRevB.88.024429, DSMcClure}. In the contrary, the GGA + $U$ calculations indicate a ferromagnetic metal with FM Ir$-$Ir coupling within the Ir$^{5+}_2$O$_9$ dimer, which is expected~\cite{PhysRevLett.116.097205} and that is 16.5 meV/f.u. lower than the AFM coupling. This demonstrates the importance of the SOC in describing the magnetic and electronic structure correctly. 

However, the generation of a local moment at the Ir$^{5+}$-$d^4$ site is complex. Ideally and owing to the large crystal field splitting of 3-4 eV, high SOC, and low pairing energy, the Ir$^{5+}$-$d^4$ should be hosting a $j_{\rm eff}$=0 state~\citep{PhysRevB.97.064408,PhysRevLett.116.097205, PhysRevLett.123.017201,annurev-conmatphys-020911-125045}. In contrast, the local moments are spontaneously generated at the Ir$^{5+}$-$d^4$ site within the GGA + $U$ + SOC calculations that are in agreement with the present experimental results. The results are similar to Ba$_3$ZnIr$_2$O$_9$ iridate, including the proposed mechanism for the Ir$^{5+}$-$d^4$ moment formation~\citep{PhysRevB.91.054412,PhysRevLett.116.097205}. The Ir$^{5+}$ moment is spontaneously generated due to the superexchange between the occupied $j_{\rm eff}$ = 3/2 and empty $j_{\rm eff}$ = 1/2 states within the Ir$^{5+}_2$O$_9$ dimer. However, in the absence of SOC, a ferromagnetic ground state emerges within the GGA + $U$ calculations, due to spin conserved hopping of electrons between the Ir$^{5+}$-$t_{2g}$ orbitals. The hopping between the half-filled Co$^{2+}$-$e_{g}^2$ and empty Ir$^{5+}$-$e_g^0$ states triggers the ferromagnetic superexchange according to the Goodenough-Kanamori-Anderson rule, consistent with the experimental magnetic structure obtained here.

\begin{figure}[t]
\centering
\includegraphics[scale=0.29]{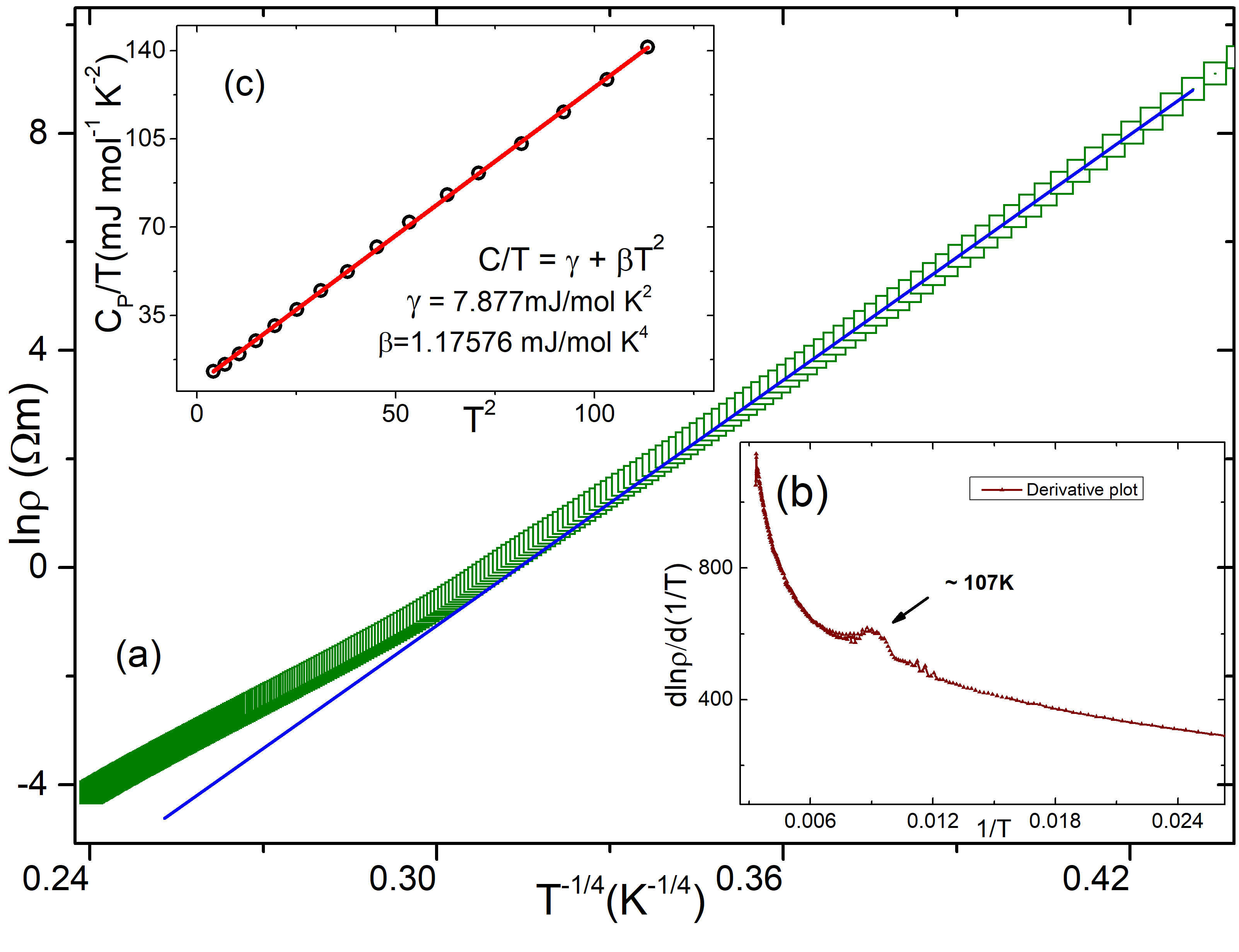}
 \caption{(a) Resistivity (ln$\rho$) plotted as a function of temperature (T$^{-1/4}$). The blue line is the linear fit to the high temperature region using the Variable Range Hopping formalism (b) Derivative of ln$\rho$ exhibits a peak close to 107 K marking the onset of the magneto-structural transition. (c) depicts the fitting to the low temperature specific heat data.}
\label{figure7}
\end{figure}

\begin{figure}[!t]
\begin{center}
{\includegraphics[width=0.47\textwidth]{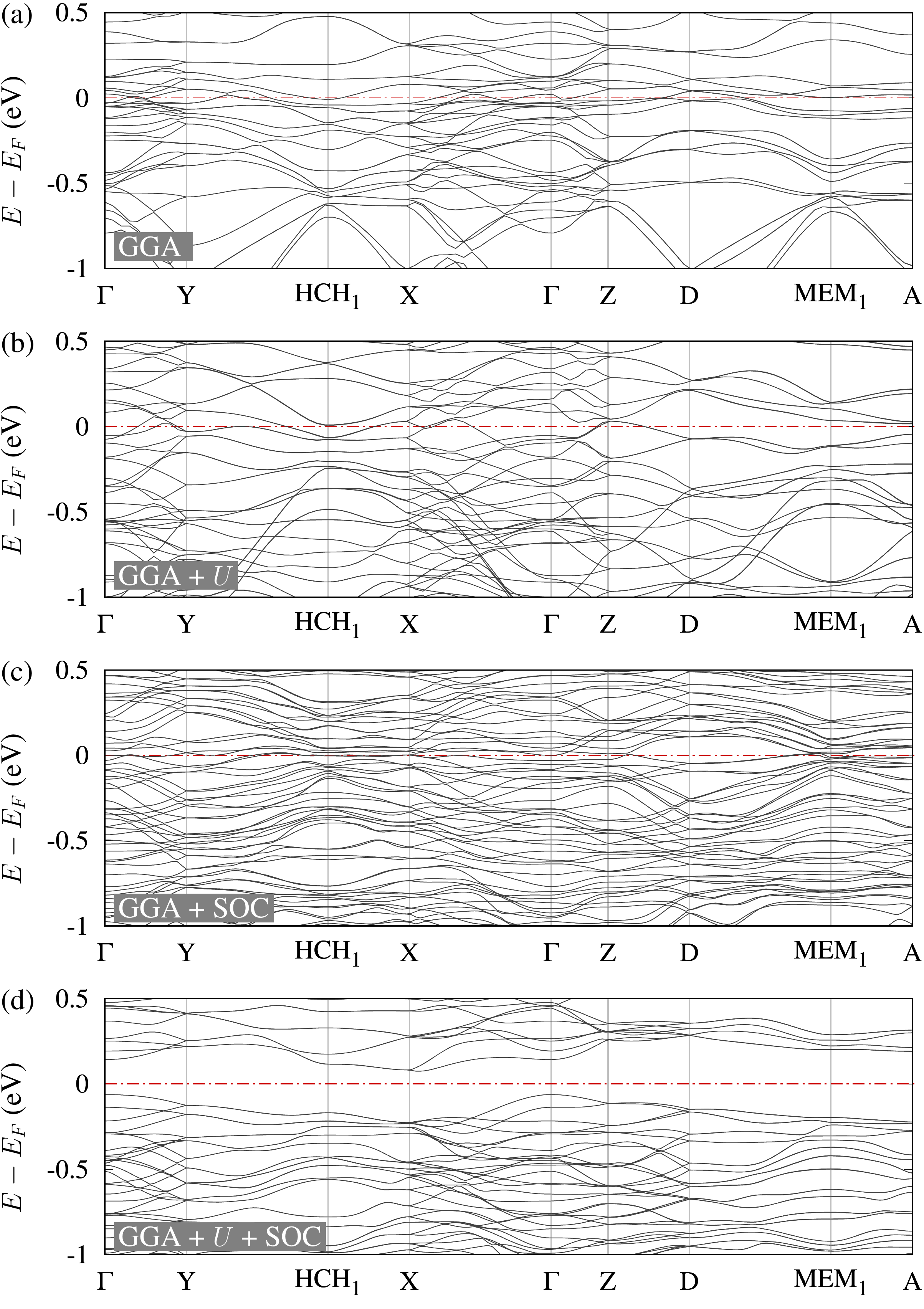}}
\caption{The calculated band structure at various theoretical hierarchy, (a) GGA, (b) GGA + $U$ ($U_{\rm Co}=$ 3 and $U_{\rm Ir}=$ 2.4 eV), (c) GGA + SOC, and (d) GGA + $U$ + SOC for Ba$_3$CoIr$_2$O$_9$. While the GGA + $U$ + SOC ($U_{\rm Co}=$ 3 and $U_{\rm Ir}=$ 2.4 eV) results correctly predicts a Mott insulating electronic structure along with the correct magnetic solution, all other calculations predict metallic band structure. These calculations together indicate the interplay between the Coulomb interaction $U$ and spin-orbit coupling. 
}
\label{fig:figure_t}
\end{center}
\end{figure}

Measurements of the temperature dependent resistivity indicate that the transport properties are governed by Variable Range Hopping (VRH) dynamics in the entire temperature range, and a change in the slope demarcates this semiconductor to semiconductor transition. Fig.~\ref{figure7} shows a plot of ln$\rho$ vs $T^{1/4}$, where the blue line is the linear fit in the high temperature region. The magneto-structural transition is seen more clearly in the derivative plot, which shows a peak at $\sim$ 107 K as shown in the lower inset. Fig.~\ref{figure7}(c) shows the low temperature specific heat data fit to the expression $  C_{p} = \gamma T + \beta T^{3} $, where $\gamma$ and $\beta$ are related to the electronic and vibrational degrees of freedom respectively. The density of states at the Fermi surface ($N\epsilon_{F}$) was deduced using the expression $ \gamma = \gamma _{0} (1+ \lambda _{e-ph}) $, where $\gamma _{0}  = \frac{\pi^{2} k_{B}^{2}}{3} N\epsilon _{F}$, where k$_{B}$ is the Boltzmann constant, and $\lambda _{e-ph}$ is the electron phonon coupling which can be ignored here as the system is insulating at low temperatures. The obtained density of states is 3.3 eV$^{-1}$f.u.$^{-1}$, which is consistent with the semi-conducting nature of Ba$_{3}$CoIr$_{2}$O$_{9}$.Though an insulating electronic state is expected due to the antiferromagnetic intradimer Ir$^{5+}$$-$Ir$^{5+}$ superexchange coupling, all the calculations excluding the GGA + $U$ + SOC predicts a metallic structure indicating the important interplay of Coulomb interaction and SO coupling (Fig.~\ref{fig:figure_t}). For example, the GGA + $U$ calculation predict a ferromagnetic metal [Fig.~\ref{fig:figure_t}(b)]. In contrast, the GGA + $U$ + SOC calculation with $U_{\rm Co}=$ 3 and $U_{\rm Ir}=$ 2.4 eV on-site Coulomb interactions correctly predicts an insulating state with 189 meV electronic gap [Fig.~\ref{fig:figure_t}(d)] that is consistent with the resistivity data. The valence and conduction bands are composed with the hybridized Ir-$d$$-$O-$p$ states, and thus the Coulomb interaction at the Ir-site is important for the insulating state. The calculated gap in Fig.~\ref{fig:figure_t}(d) depends on $U_{\rm Ir}$ and decreases to 136 and 60 meV for  $U_{\rm Ir} = $ 2 and 1.5 eV, respectively, below which the system becomes metallic.

\begin{figure}
\centering
\includegraphics[scale=0.28]{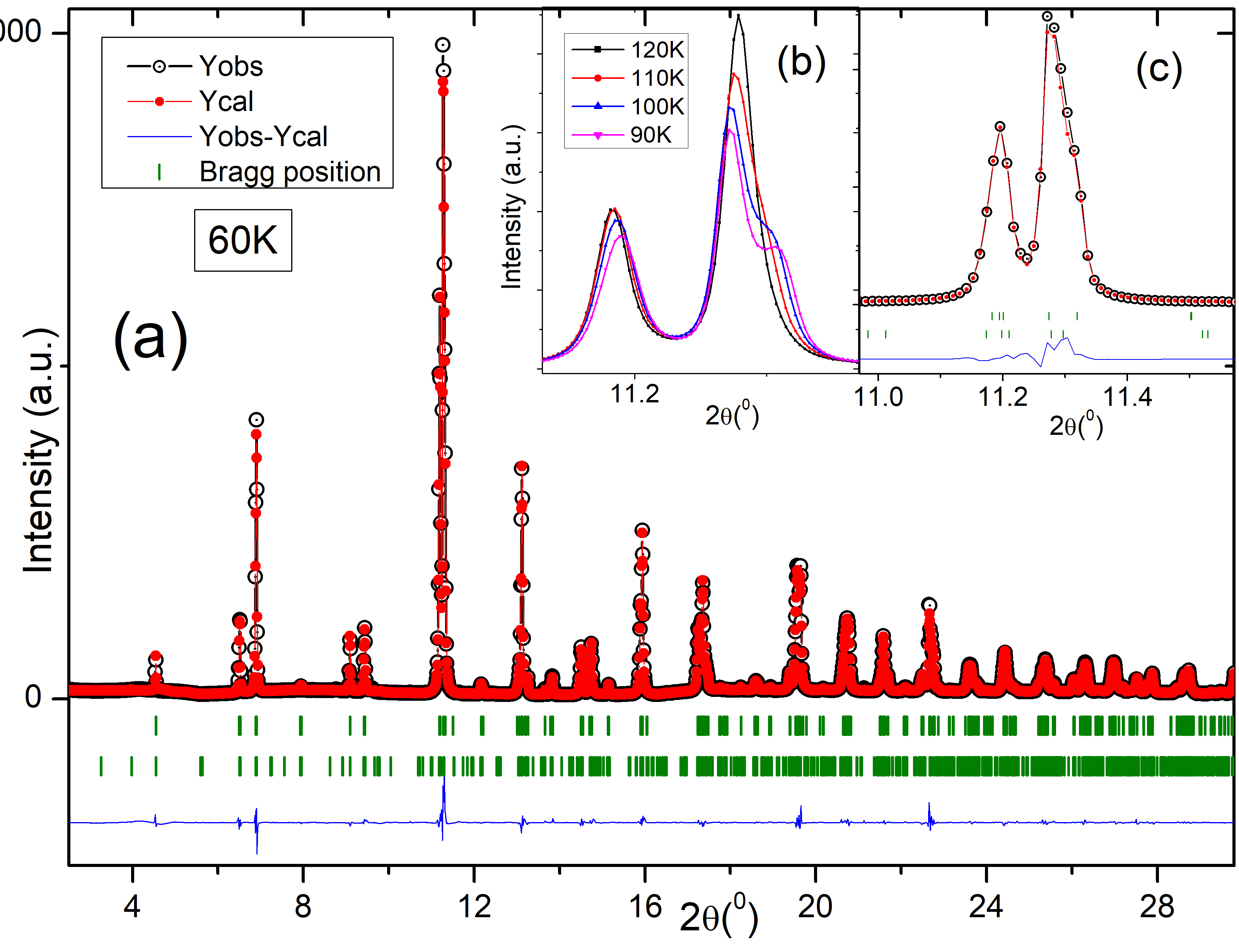}
\caption{Main Panel: Rietveld refinement of synchrotron powder diffraction pattern of Ba$_{3}$CoIr$_{2}$O$_{9}$ at $60\,$K. The calculated and observed diffraction profiles are shown as red and black markers respectively. The vertical green lines indicate the calculated Bragg positions for the monoclinic  $C2/c$ (top) and $P2/c$ (bottom) phases. The blue line at the bottom depicts the difference between observed and calculated intensities. The inset (b) depicts the temperature dependence of the main peaks, with the onset of the magnetostructural transition seen below 110\,K. The inset (c) shows an expanded view of the main peak fitted using the two-phase model at 60\,K. }
\label{figure9}
\end{figure}
	
\begin{figure}[!t]
\centering
\includegraphics[scale=0.36]{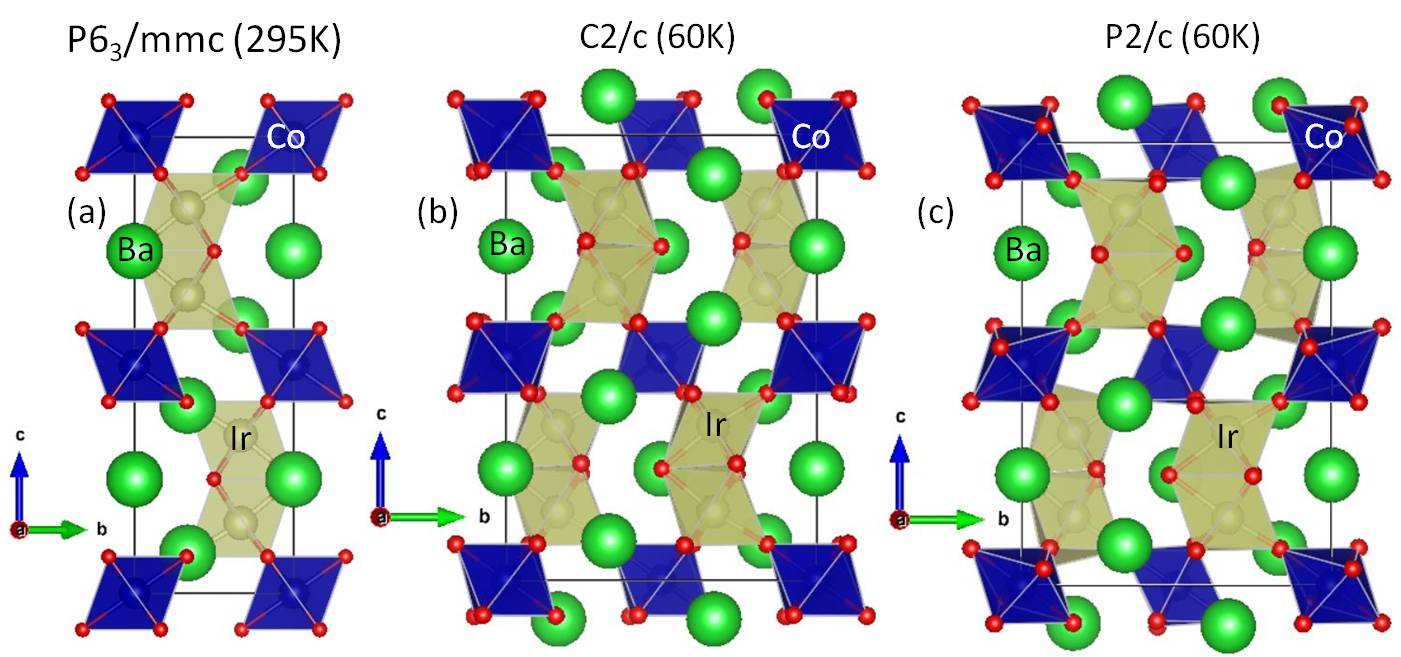}
\caption{Main Panel: Schematic crystal structures exhibited by Ba$_{3}$CoIr$_{2}$O$_{9}$ across the temperature range of our investigations.  Here, the blue and yellow octahedra represent the cobalt and iridium sites respectively. The high symmetry $P6{_3}/mmc$ structure (a) transforms to a monoclinically distorted $C2/c$ structure (b) at the magnetostructural transition. On further reduction in temperature, a part of the system transforms to an even lower symmetry $P2/c$ (c), with both these monoclinic phases coexisting down to the lowest measured temperatures.}
\label{figure10}
\end{figure}

The fitting of the low temperature XRD data reveals a gradual distortion of the structure to an even lower monoclinic symmetry ($P2/c$) below 70K, which coexists with the monoclinic $C2/c$ phase down to the lowest measured temperature. The Rietveld refinement of the diffraction pattern measured at 60 K, which has been fit using coexisting $C2/c$ and $P2/c$ monoclinic phases is shown in Fig.~\ref{figure9}. The inset shows an expanded view of the fit of the major Bragg peak where the incorporation of both the phases can be seen explicitly. The refinement parameters R$_{p}$, R$_{wp}$ and R$_{e}$ obtained for this fit was 6.4, 9.9 and 1.2 respectively, and all our low temperature fits yielded similar values. The crystal structure of Ba$_{3}$CoIr$_{2}$O$_{9}$ at room temperature, as well as the two monoclinic phases at 60 K is depicted in Fig.~\ref{figure10}, and the structural parameters for these phases (at 60 K), as deduced from the refinement procedure, are summarized in Table~\ref{tab:table2}.  

\begin{table}[!b]
\centering
\caption{Structural parameters of Ba$_{3}$CoIr$_{2}$O$_{9}$ at 60 K.}
\begin{tabular}{C{1.5cm} C{1.7cm}  C{1.8cm}  C{1.5cm} C{1.5cm}}
\hline
\hline
\multicolumn{5}{c}{Space group $C2/c$ (No. 15)} \\
\multicolumn{5}{c}{$a = 5.7340(1)$ \AA, $b = 9.9841(1)$ \AA, $c = 14.2652(2)$ \AA} \\
\multicolumn{5}{c}{$\alpha=\gamma$ = 90$^{\degree}$, $\beta$ = 90.0385(10)$^{\degree}$} \\
\hline
Atom & $x$  & $y$ & $z$ & occupancy\\ \hline
    Ba1 & 0.0004(13) & 0.3312(05) & 0.0877(03) & 1.00\\ 
    Ba2 & 0.0000 & -0.0006(07) & 0.2500 & 1.00\\
    Co & 0.0000 & 0.0000 & 0.0000 & 1.00\\
    Ir & -0.0060(08) & 0.3337(04) & 0.8462(05) & 1.00\\
    O1 & 0.0000 & 0.5030(29) & 0.2500 & 1.00\\
    O2 & 0.2662(62) & 0.2414(30) & 0.2381(18) & 1.00\\
    O3 & 0.0296(70) & 0.8401(29) & 0.0805(33) & 1.00\\
    O4 & 0.2567(51) & 0.1003(24) & 0.0814(19) & 1.00\\
    O5 & 0.7709(83) & 0.0744(56) & 0.0784(50) & 1.00\\
\hline
\multicolumn{5}{c}{Space group $P2/c$ (No. 13)} \\
\multicolumn{5}{c}{$a = 5.7550(1)$ \AA, $b = 9.9452(1)$ \AA, $c = 14.2634(2)$ \AA} \\
\multicolumn{5}{c}{$\alpha = \gamma$ = 90$^{\degree}$, $\beta$ = 90.2291(19)$^{\degree}$} \\
\hline
Atom & $x$  & $y$ & $z$ & occupancy\\ \hline
    Ba1 & 0.5049(22) & 0.1660(19) & 0.9156(08) & 1.00\\ 
    Ba2 & 0.0000 & 0.0017(22) & 0.2500 & 1.00\\
    Ba3 & 0.5000 & 0.4986(27) & 0.2500 & 1.00\\
    Ba4 &-0.0018(23) & 0.6676(18) & 0.9092(07) & 1.00\\
    Co1 & 0.0000 & 0.0000 & 0.0000 & 1.00\\
    Co2 & 0.5000 & 0.5000 & 0.0000 & 1.00\\
    Ir1 & 0.4989(17) & 0.1671(13) & 0.1554(06) & 1.00\\
    Ir2 & -0.0015(18) & 0.6654(13) & 0.1532(07) & 1.00\\
    O1 & 0.729(19) & 0.241(11) & 0.263(44) & 1.00\\
    O2 & 0.180(15) & 0.748(10) & 0.247(53) & 1.00\\
    O3 & 0.0000 & 0.521(09) & 0.7500 & 1.00\\
    O4 & 0.5000 & -0.008(21) & 0.7500 & 1.00\\
    O5 & 0.277(19) & 0.089(14) & 0.4142(55) & 1.00\\
    O6 & 0.799(20) & 0.5636(79) & 0.4209(58) & 1.00\\
    O7 & 0.019(18) & 0.165(13) & 0.9162(75) & 1.00\\
    O8 & 0.482(23) & 0.674(12) & 0.919(10) & 1.00\\
    O9 & 0.728(16) & 0.071(11) & 0.4618(43) & 1.00\\
    O10 & 0.267(16) & 0.562(14) & 0.4077(39) & 1.00\\
\hline
\end{tabular}
\label{tab:table2}
\end{table}

\begin{figure}[t]
\centering
\includegraphics[scale=0.32]{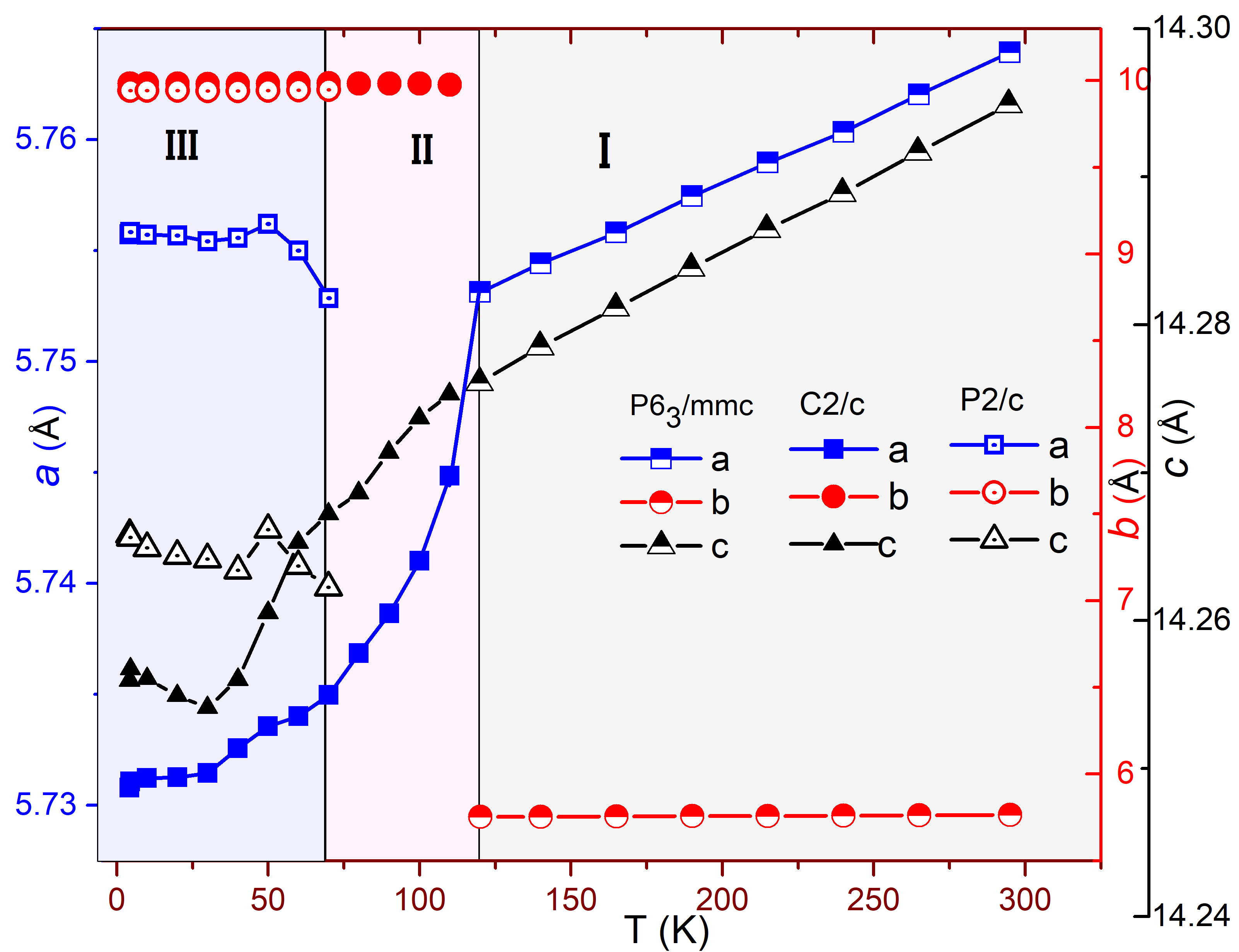}
\caption{Variation of the lattice parameters of Ba$_{3}$CoIr$_{2}$O$_{9}$ as a function of temperature. The error bars are of the order of the symbol sizes. The three regions correspond to three  different structural phases. Region I - $P6_{3}/mmc$, II - $C2/c$ and III -$C2/c+P2/c$. }
\label{figure11}
\end{figure}

The temperature dependence of the lattice parameters as determined in Ba$_{3}$CoIr$_{2}$O$_{9}$ is depicted in Fig.~\ref{figure11}. The lattice parameters $a$ and $c$ decrease on lowering the temperature, whereas the lattice parameter $b$ increases sharply across the magneto-structural transition. The ensuing distortion in the lattice is also evidenced in the form of octahedral tilting and distortion, as is also corroborated by considerable changes in the Co-O and Ir-O bond lengths and angles.  The phase fractions of the $P2/c$ and $C2/c$ phase as a function of temperature is depicted in the inset of Fig.12, and the $P2/c$ phase is observed to increase at the cost of the $C2/c$ phase with decreasing temperatures. However, at around 30K, this trend is seen to reverse. We note that a similar co-existence (and competition) between these two phases is been reported earlier in the vanadate perovskites of the form $R$VO${_3}$ (with $R$ = Sm, Ho, Yb, Pr or Y)~\cite{RVO}. We believe that this is the first report of such phase coexistence in the triple perovskite iridates, and it remains to be seen whether other members of this extended family also exhibit similar features.
 
\begin{figure}[t]
\centering
\includegraphics[scale=0.27]{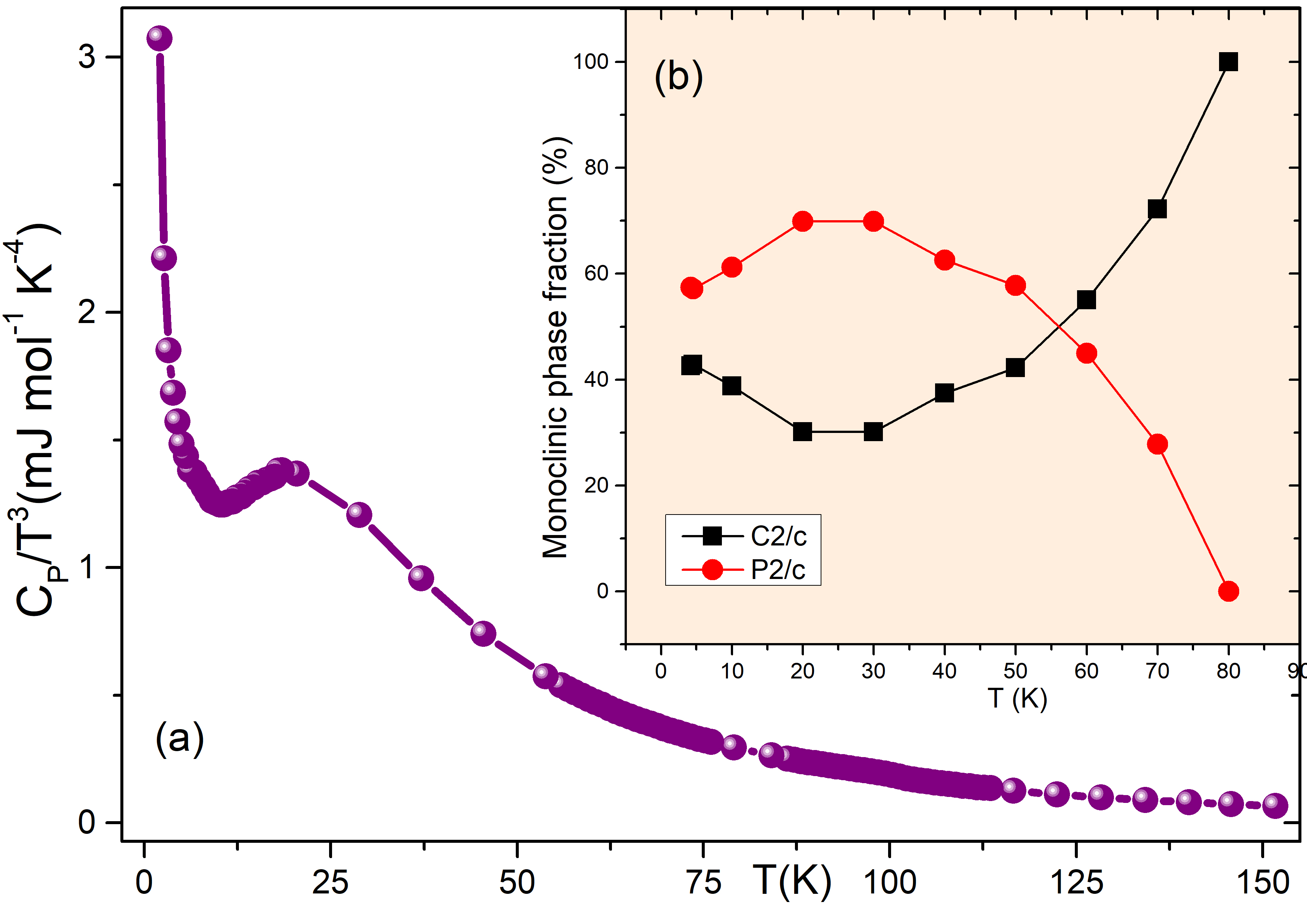}
\caption{The main panel depicts the low temperature specific heat $C_p/T^3$ as a function of temperature as measured in Ba$_{3}$CoIr$_{2}$O$_{9}$. The inset depicts the temperature evolution of the two monoclinic phases as determined from the analysis of structural data, with the uncertainty being of the order of the symbol sizes. }
\label{figure12}
\end{figure}

The co-existence of two structurally disparate phases in Ba$_{3}$CoIr$_{2}$O$_{9}$ also manifests itself in the form of a pronounced deviation of the measured specific heat from the standard Debye $T{^3}$ dependence at low temperatures. This is seen in the main panel of Fig.~\ref{figure12}, where $C{_P}/T{^3}$ is plotted as a function of temperature. This excess entropy is a signature of a glassy phase~\cite{Cp} and is similar to what is seen in structural glasses, as well as in systems like the mixed valent manganites, where the electronic phase separation results in the formation of magnetically (and structurally) dissimilar phases which co-exist with each other.  In most of the manganites, these competing phases are ferromagnetic and antiferromagnetic \cite{Cp1,Cp2,Cp3}, whereas in this case, both the phases appear to be predominantly antiferromagnetic, albeit with varying amounts of monoclinic distortion. 

\begin{figure}[t]
\centering 	
\includegraphics[scale=0.35]{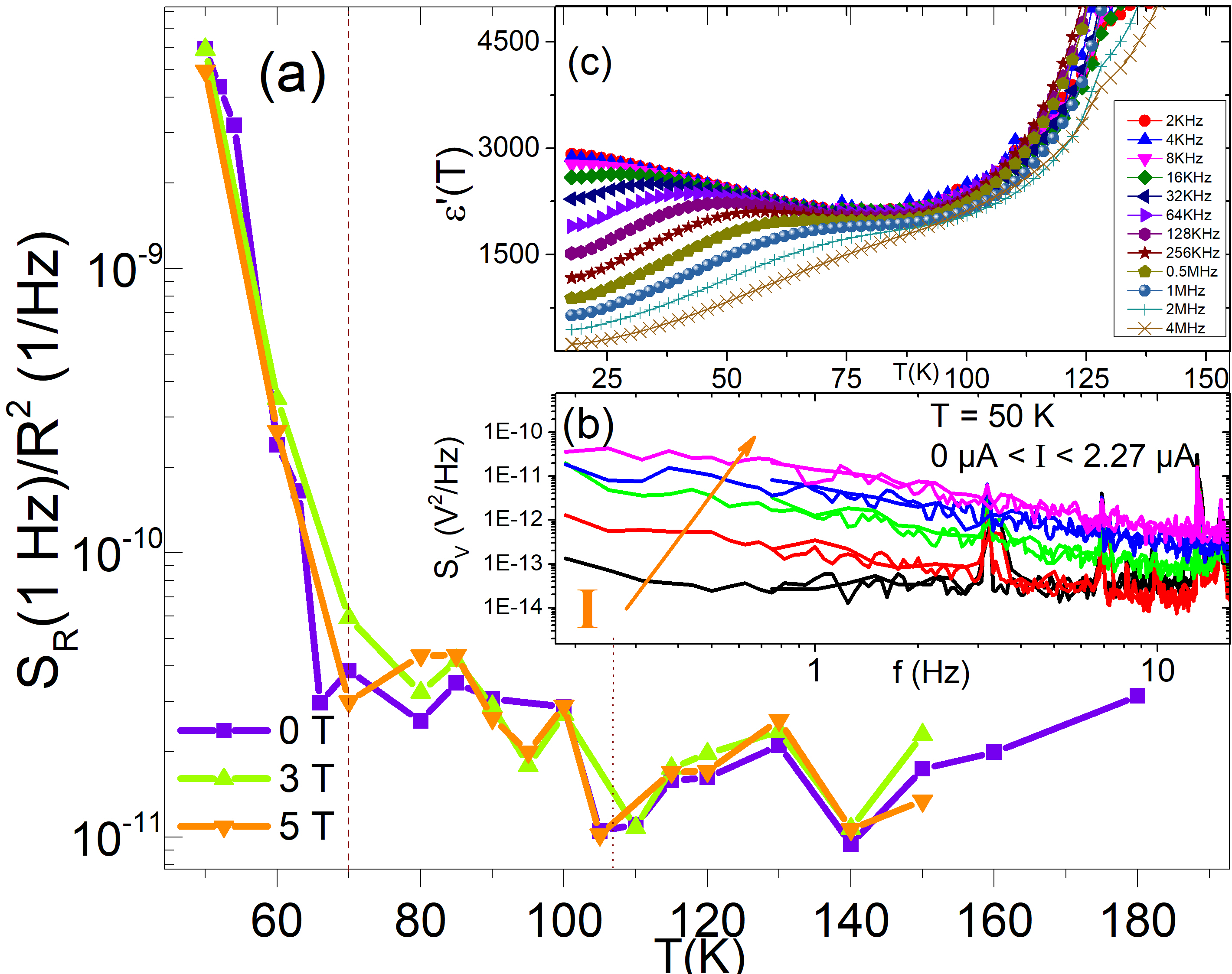}
\caption{The main panel depicts the temperature-dependent normalized resistance noise evaluated at $1\,$Hz for Ba$_{3}$CoIr$_{2}$O$_{9}$ at different magnetic fields. The inset (b) depicts the scaling of the voltage noise magnitude $S_V$ with the applied current $I$ at a fixed temperature of $50\,$K. (c) depicts the temperature dependence of the real part of the dielectric susceptibility as measured in frequencies ranging from $2\,$KHz to $4\,$MHz. } 
\label{figure13}
\end{figure}

We have also investigated the Ba$_{3}$CoIr$_{2}$O$_{9}$ using resistance noise spectroscopy, which provides an interesting perspective - especially near phase transitions - in strongly correlated oxides. Fig.~\ref{figure13}(a) shows the normalized resistance noise PSD evaluated at f = 1 Hz.  Noise measurements were performed at temperatures between $50\,$K and $180\,$K and for various external magnetic fields between $0\,$T and $7\,$T.  In order to verify that the measured $1/f$-type noise spectra originate from the investigated sample, we checked the scaling of the voltage noise magnitude $S_V$ with the applied current $I$. After Hooge's empirical law ~\cite{HOOGE1969139, HOOGE197614}, 
\begin{equation}
S_V(f)=\frac{\gamma_{\mathrm{H}}V^2}{n\Omega f^{\alpha}},
\end{equation}   
the magnitude of the voltage noise is expected to scale as $S_V\propto V^2 \propto I^2$. Here, $V$ represents the applied voltage, $\gamma_{\mathrm{H}}$ the material-dependent Hooge parameter, $\alpha$ the frequency exponent, $n$ the charge carrier density, and $\Omega$ the "noisy" sample volume, i.e.\ $n\Omega=N_c$ gives the total number of charge carriers in the material causing the observed $1/f$ noise. The inset(b) of Fig.~\ref{figure13} demonstrates the expected scaling behavior of the acquired noise spectra at a selected temperature of 50 K. Note that for zero applied current, only the frequency-independent background noise of the experimental setup is observed. As depicted in the main panel of Fig.~\ref{figure13}, the normalized noise magnitude exhibits only a weak temperature dependence at high temperatures. At the onset of the magneto-structural transition, the noise is observed to increase by almost an order of magnitude. Moreover, on further reduction in temperature, a drastic increase of about two orders of magnitude is observed between $70\,$K and $50\,$K, where our structural data infers on the co-existence of the two structurally disparate monoclinic phases.  There is no pronounced and systematic dependence of the normalized noise magnitude on the applied magnetic field. Even though there is no additional signature in the resistivity below the phase transition, the resistance noise is clearly far more sensitive to the onset of phase coexistence than to the magnetostructural transition at $107\,$K. A possible explanation for the strongly enhanced $1/f$-type fluctuations below $70\,$K is the presence of inhomogeneous current paths due to the occurrence of different structural phases. These coexisting phases can be assumed to differ in their individual electronic properties. It is well known for other materials with coexisting electronic phases that an inhomogeneous conduction can lead to strong modifications of the measured noise magnitude ~\cite{PhysRevB.97.054413, cryst8040166}. 

The inferences drawn from the noise measurements are also reinforced by our dielectric data, with the real part of the dielectric susceptibility exhibiting a substantial frequency dependence in the magnetically ordered state, with this dispersion being even more pronounced below $70\,$K where we have coexisting monoclinic phases (Fig.~\ref{figure13}c). The fact that we observe this frequency dependence only below the magnetic transition temperature clearly indicates a coupling of electric and magnetic order parameters. We note that such a frequency dependent feature in dielectric susceptibility is characteristic of charge relaxation processes, and has been observed in a number of phase separated systems~\cite{PhysRevB.72.144429}. However, our attempts to fit the frequency dependent peaks to both the Arrhenius and Vogel-Fulcher-Tammann (VFT) formulations were unsuccessful, indicating that the dynamics of these charge relaxations and also their magnetic field dependence would need additional investigations. 
	
\section{Summary}
In summary, we report on a comprehensive experimental and theoretical investigation of the triple perovskite iridate Ba$_{3}$CoIr$_{2}$O$_{9}$. Stabilizing in the 
hexagonal $P6_{3}/mmc$ symmetry at room temperature, it exhibits a magneto-structural transition to a monoclinic $C2/c$ phase at 107 K -- the highest known amongst all the triple perovskite iridates. Below 70 K, a part of the system transforms to a monoclinic phase with even lower symmetry ($P2/c$), and both these phases coexist down to the lowest measured temperatures. First-principles calculations, including the spin-orbit coupling within the GGA + $U$ + SOC formalism, predicts a weakly b-canted AFM structure as the magnetic ground state for the $C2/c$ phase that agrees well with the magnetic structure derived from the neutron diffraction data. The observation of excess entropy in the low temperature specific heat, a pronounced increase in the resistance noise and the frequency dependent dispersion in the dielectric susceptibility, all point towards a highly correlated ground state in Ba$_{3}$CoIr$_{2}$O$_{9}$.  
 
\section{Acknowledgements}
C.G and S.N. acknowledge Surjeet Singh for extending experimental facilities, and Sugata Ray for help in analyzing XPS data.  C.G., M.L., J.M. and S.N. acknowledge DST India for support through grant no. INT/FRG/DAAD/P-249/2015. M. K. and S. N. acknowledge the DST Nanomission Thematic Unit Program, SR/NM/TP-13/2016. C.G. and S.N. thank the Department of Science and Technology, India (SR/NM/Z-07/2015) for  financial support and Jawaharlal Nehru Centre for Advanced Scientific Research (JNCASR) for managing the project. D. R. is grateful to the Council of Scientific and Industrial Research (CSIR), Government of India for financial support in the form of a research fellowship.
  
\bibliography{Bibliography}

\end{document}